# Surface conductivity of DR1-functionalized organic-inorganic sol-gel materials.+

J.-A. Reyes-Esqueda[1], A. Franco[1], M. Bizarro[1], J. García-Macedo[1], M. Canva[2], B. Darracq[2], Y. Lévy[2], K. Lahlil[3], F. Chaput[3], J-P. Boilot[3].

[1] Departamento de Estado Sólido, Instituto de Física – UNAM, Ciudad Universitaria, Del. Coyoacán. 04510, México, D.F., México. E-mail: reyes@fenix.fisica.unam.mx

[2] Laboratoire Charles Fabry de l'Institut d'Optique.
Centre National de la Recherche Scientifique – CNRS UMR 8501
Université d'Orsay-Paris XI, BP 147, 91403 Orsay Cedex, France

[3] Laboratoire de Physique de la Matière Condensée.
Centre National de la Recherche Scientifique – CNRS UMR 7643
École Polytechnique, 91128 Palaiseau, France

**Abstract.**

In recent years, in looking for an important photorefractive application, several studies on electro-optic and photoconductive sol-gel responses have been done. A very important effort has been addressed to establish the appropriate induced-orientation procedure, in order to get the highest electro-optic coefficient. In this way a very high coefficient of 48 pm/V at 831 nm in sol-gel has been already found. Similarly, the importance of the non-linear chromophore concentration into the material electro-optic behavior has been studied. However, the influence of the orientation procedure and the chromophore concentration over the photoconductive response has not been performed. In this work we study the vacuum-surface-charge-transport under and without illumination after poling times of 10, 30 and 120 min on DR1-functionalized sol-gel thin films of 1.3 μm in thickness with a suitable concentration of DR1. We include the measures before poling for other chromophore concentrations. We found the largest density of photocurrent at 633 nm for a poling time of 30 min. We also measured the order parameter in order to follow the Corona induced orientation evolution as function of time for each case. The saturation found into this parameter and into the photoconduction show the existence of an optimal poling time.

Keywords: sol-gel, photorefractive, photoconduction, electro-optic, chromophore concentration, molecular orientation,...

+: Work supported by Conacyt 34582E and DGAPA UNAM IN 103199, Mexico, and by the Ministère de la Recherche, France.

## 1. Introduction.

Recently, several studies have been performed in all kind of amorphous materials in order to get a photorefractive response[1,2,3]. The interest has been mainly focused in polymers due to their large induced-nonlinearities, low dielectric constant, structural flexibility and low cost[4]. This effort has also been addressed to sol-gel materials[5,6]. On them, a very important work has been done to establish the appropriate induced-orientation procedure, in order to get the highest electro-optic coefficient[7,8]. At the same time, there have been found several important results of the non-linear chromophore concentration influence on the material electro-optic behavior[9,10]. However, the influence of both, the orientation procedure and the chromophore concentration over the photoconductive response, has not been performed. In this way, some results show that molecular orientation is decisive into the photocurrent measurements[11]. Takimoto et. al. show how thin films with parallel-oriented oligo-p-phenylene sulfide (OPS) molecules can be obtained by vapor deposition, and how these molecules have anisotropic photoconduction, that is, carriers are easier transported perpendicular to the molecular chain axis than along that axis[11]. In their work they propose the next model to explain this result. They say that carrier transport occurs by means of an interchain hopping between oligomers. OPS molecules consist of zigzag chains of sulfur atoms and benzene rings. The benzene rings of the neighboring chains have the same direction, and π orbitals overlap each other to form carrier paths perpendicular to the chains as shown in figure 1. There exist some references about anisotropic carrier transport in organic materials where carriers move easily along the stacking direction of the aromatic rings[12,13]. In some cases, this stacking direction coincides with the direction perpendicular to the chain axis. Therefore, carriers move easily perpendicular to the chain axis[13].

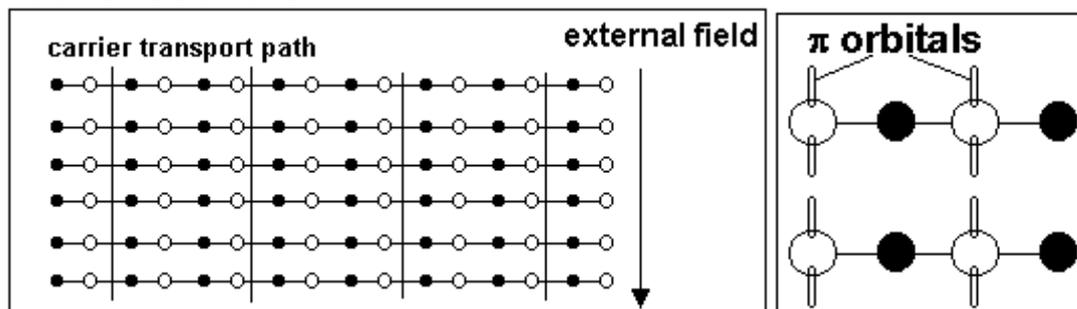

Figure 1. Schematic diagram of carrier transport model for OPS molecules. (●) sulfur atom, (○) benzene ring.

In this work, the sol-gel method was used to prepare hybrid organic-inorganic thin films with the classical non-linear optical (NLO) azo-chromophore DR1 (disperse red one) grafted into the skeleton matrix. The DR1 concentrations with respect to the carbazole molecule (K) were of SiDR1:SiK=1:2, 1:5 and 1:20. Carbazole is a charge-transporting unit and it could also be attached to the sol-gel silicon backbone. We will show how, by taking the model explained above, the vacuum-surface-charge-transport in general, and the surface photoconduction at 633 nm in particular, grew as long as the DR1 molecules were oriented by the Corona technique for the 1:5 concentration. We also measured the order parameter in order to follow the evolution on time of the Corona induced order for each case. The order parameter was calculated according to[14]

$$\rho = 1 - \frac{A_\perp}{A_0} \qquad (1),$$

where $A_\perp$ ($A_0$) is normal absorption after (before) poling. This equation is valid for a rod-like molecule and for a homogeneous chromophore concentration as it is supposed in this work.

**2. Experimental.**

The sample preparation has been previously reported[5]. The molecular precursors were prepared by reacting carbazole-9-carbonyle chloride with 3-aminopropyltriethoxysilane, yielding the charge-transport precursor SiK, and by reacting the chromophore bearing free hydroxyl group with 3-isocyanatopropyltriethoxysilane yielding the push-pull NLO precursor SiDR1. To prepare coating solutions, the modified silanes in this way were copolymerized with tetraethoxysilane (TEOS) used as cross linking agent as follows: The hydrolysis of these groups was made by adding acidulated water (pH=1) to the solution using tetrahidrofuran (THF) as a common solvent. The molar ratio for this was (SiDR1+SiK+TEOS):H$_2$O:THF=1:28:47. After 2 hrs of hydrolysis, a little amount of pyridine was incorporated in order to neutralize the acidity of the solution and then to promote the condensation reactions. The solution was filtered (45 μm) before depositing the samples by spin-coating. The NLO chromophore concentration was modulated with respect to the SiK concentration. The relative composition to the TEOS agent was fixed as (SiDR1+SiK):TEOS=6:1, and the SiDR1:SiK studied compositions were of 1:2, 1:5 and 1:20. Spin coating on glass substrates made films with thickness around of 1.3 μm. After the preparation, the films were dried at 120 °C during 1 hr under vacuum, and then the solvent was evaporated by a thermal treatment at 120 °C during 15 hrs. After poling, silver electrodes with a length around of 3 mm were painted on the surface of the film in order to measure conductivity.

The linear absorption was measured using a Milton Roy spectrophotometer model Spectronic 2PC. The orientation of the DR1 molecules in the sol-gel film (and then the nonlinear optical properties of the material[10]) was induced by the single-point Corona poling technique (needle-surface distance=12 mm, voltage=+6 kV, poling temperature=120 °C). The poling times studied were 10, 30 and 120 min. It is necessary to say that a new, different sample was used for every poling time studied. In this way, for each of these times, the separation length of the painted electrodes was $d$=0.34, 0.39 and 0.24 cm, respectively. The vacuum-surface-conductivity was measured after each poling process according to the experimental scheme shown in figure 2. Picocurrents were measured into the 0-500 V interval, with steps of 100 V. The photocurrent was generated by using an Oriel 79309 10 mW He-Ne laser. The absorption was measured before and after every poling process. For the non-oriented samples the same photoconduction measuring process was followed.

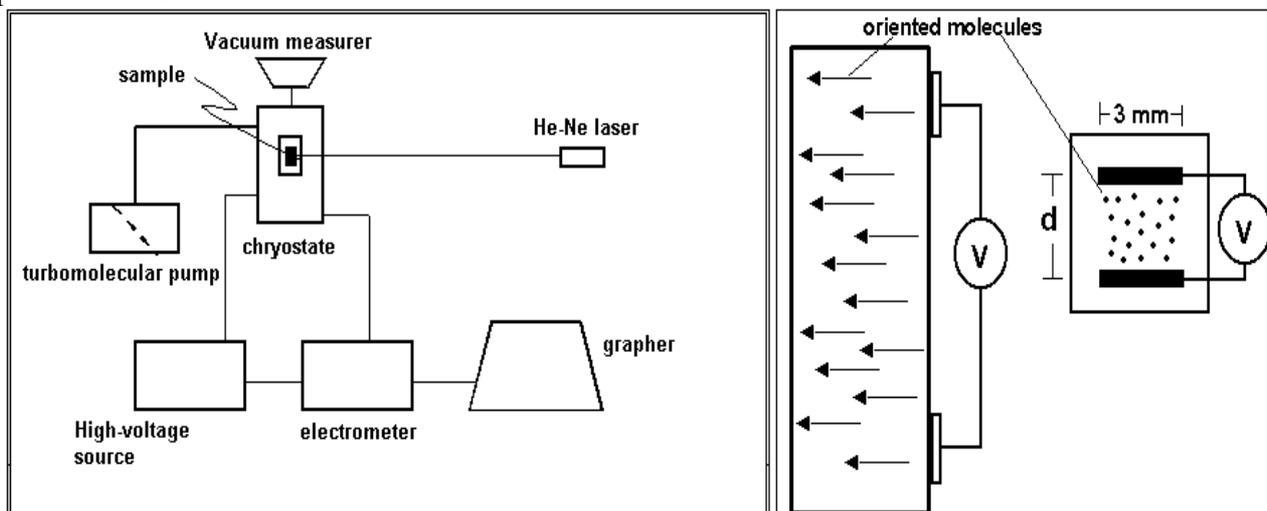

Figure 2. Experimental scheme for conductivity measurements. Lateral and frontal view of the sample inside the cryostat for an idealized 100 % molecular orientation.

**3. Results and discussion.**

The molecular orientation induced in these samples is corroborated by the increment in the order parameter as a function of poling time, as it is shown in figure 3 for the SiDR1:SiK=1:5 concentration (where a extra poling time of 1 hr is included). This curve can be well fitted by a Langevin-Debye equation (black line). The experimental points show saturation at 90 % for the fitting curve after around 1 hr of poling, indicating that there is an asymptotical polarization for the molecules.

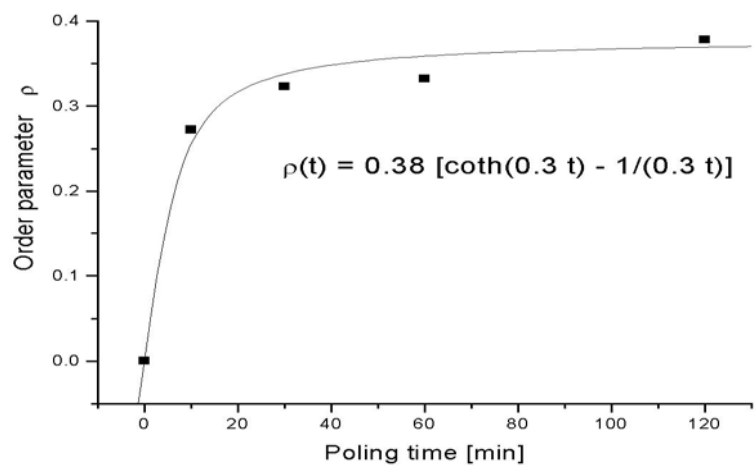

Figure 3. Order parameter as a function of poling time.

The current density was calculated by taking the transversal area as the product of the electrode average length (3 mm) and the film thickness (1.3 μm). The applied field was calculated by taking the distance between the electrodes, *d*. For the non-oriented samples the currents were the same, i.e., very small, for the three concentrations studied, as it is shown in figure 4. This fact allows seeing the importance of molecular orientation in order to get efficient charge transport.

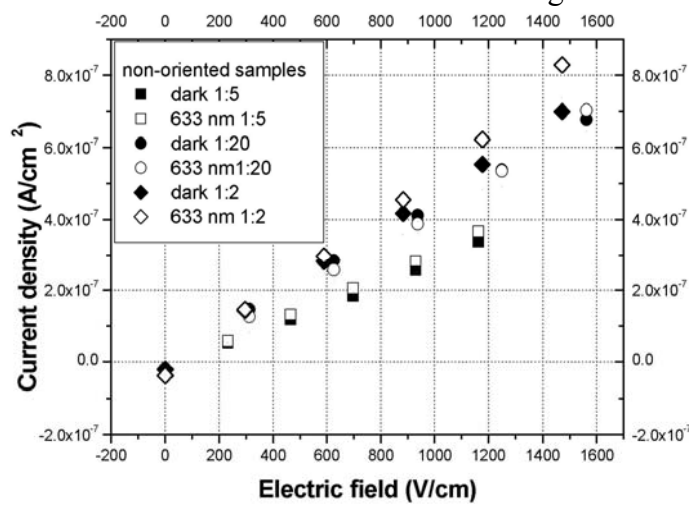

Figure 4. Current density for the non-oriented samples at SiDR1:SiK=1:2, 1:5 and 1:20 concentrations.

The results for dark- and photo-current densities for the SiDR1:SiK=1:5 concentration are shown in figure 5. There can be seen how initially the slope increases, and therefore the current density, as a function of the poling time. After 30 min there is a slight decrement into the currents. Since the DR1 molecular structure comprises two benzene rings, the results can be explained according to the model shown above by recognizing that, as DR1 molecules become oriented into the sol-gel film, perpendicular to the surface of the sample, the π orbitals of their benzene rings overlap more, and then the carrier transport of π electrons increases in the perpendicular direction to the chromophore axis, (*i.e.*) parallel to the surface of the thin film. We need to take into account the presence of the carbazole molecules, which are supposed to transport charge easier than DR1. It can be said that, besides of acting as screeners for the electrostatic interactions between non-linear chromophores[10], these molecules facilitate the overlap of the DR1 π orbitals by also overlapping theirs with them, and in this way the surface conductivity is in general increased until a minimum distance is reached. The decrement may be explained by thinking that there is a shorter distance between the oriented chromophores at 30 min of poling. This fact may imply the existence of an optimal poling time around 30 min. However, further experiments must be done to put this clear.

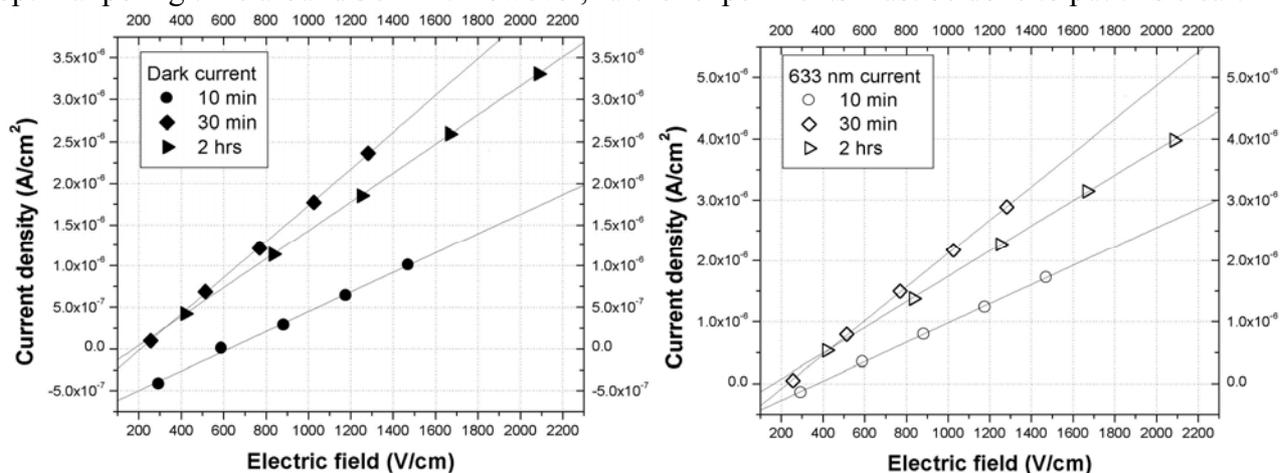

Figure 5. Current density as a function of the electric field for the SiDR1:SiK=1:5 concentration. A linear fit is applied for each curve.

A linear fit (shown in figure 5) for both, the dark- and the photo-current densities, allow to obtain the parameters given in table 1 for the oriented concentration SiDR1:SiK=1:5, according to the next equation:

$$J = \frac{e\alpha I}{h\nu}[\phi l_0 + \phi\mu\tau E] + e n_0 \mu E, \qquad (2)$$

where the first term is due to the photovoltaic effect, the second to the photoconductivity, and the third one to the dark conductivity. In this equation $e$ is the electron charge, $h\nu$ is the photon energy, $\alpha$ is the material absorption coefficient, $I$ is the light intensity, $\phi$ is the quantum efficiency to excite a free carrier, $l_0$ is the effective electronic mean free way, $\mu$ is the carriers mobility, $\tau$ is the excited carriers lifetime, $n_0$ is the dark carriers density, and $E$ is the applied field. The obtained parameters are still lower than them reported for photorefractive crystals as $KNbO_3:Fe^{3+}$.

| 633 nm | $KNbO_3:Fe^{3+}$ | | DR1 | |
|---|---|---|---|---|
| $t_{poling}$ (min) | $\phi l_0$ ($10^{-8}$ cm) | $\phi\mu\tau$ ($10^{-11}$ cm$^2$/V) | $\phi l_0$ ($10^{-12}$ cm) | $\phi\mu\tau$ ($10^{-14}$ cm$^2$/V) |
| 0 | ~0.85 | 23.38 | - | - |
| 10 | - | - | 6.06 | 1.53 |
| 30 | - | - | -11.94 | 3.82 |
| 120 | - | - | -2.05 | 2.05 |

Table 1. Sample parameters obtained from the linear fit made by taking equation (2). Known crystal parameters are shown for comparison.

## 4. Conclusions.

All these results in our SiDR1:SiK sol-gel thin films show that, when there is no molecular orientation at all, the conduction is almost zero, and then, as long as the DR1 molecules become oriented perpendicular to the surface of the thin film, the interchange of π electrons between benzene rings of carbazole and DR1 units increases in a perpendicular direction to the axis of the chromophore as poling time does, until a minimum distance between them is reached. By the other side, we can establish that our poling process is optimized for a time around 30 min of molecular orientation and that this orientation can be fitted by a usual Langevin-Debye equation. However, this result is no coincident with that found for electro-optics, where an optimal poling time of 2 hrs was found[7,8,10]. We may conclude that further experiments with oriented samples at other concentrations have to be done in order to establish a compromise between the photoconductive and the electro-optic responses in this kind of materials.

## 5. References.